%% file: ms_main.tex
\documentclass[runningheads]{llncs}
\usepackage[utf8]{inputenc}
\usepackage{graphicx}
\usepackage{textcomp}
\usepackage{booktabs}
\usepackage{amsmath}
\usepackage{amsfonts}
\usepackage{algorithm}
\usepackage{algpseudocode}
\usepackage{csquotes}
\usepackage{enumerate}
\usepackage{multirow}
\usepackage{multicol}
\usepackage{tabularx}
\usepackage{array}
\usepackage{xcolor}
\usepackage{subcaption}
\usepackage{flushend}
\usepackage{comment}
\usepackage{mathrsfs}
\usepackage{gensymb}
\usepackage{tikz}
\usepackage{cite}
\usetikzlibrary{trees}
\usepackage{dsfont}
\usepackage{mathtools}
\usepackage{listings}
\usepackage{siunitx}
\usepackage{hyperref} 
\hypersetup{
    colorlinks=true,
    linkcolor=red,
    filecolor=magenta,      
    urlcolor=red,
    citecolor=red,
    }

\lstdefinestyle{myStyle}{
    breaklines=true,
    frame=lines,
    numbers=none,
    captionpos=b,
    basicstyle=\scriptsize\ttfamily,
    keywordstyle=\bfseries\color{green!60!black},
    commentstyle=\itshape\color{purple!50!black},
    identifierstyle=\color{blue},
    backgroundcolor=\color{gray!5!white},
    escapeinside={<}{>}
}
\def \eg{\emph{e.g.}}
\def \ie{\emph{i.e.}}
\def \cf{\emph{cf.}}
\def \etal{\textit{et al.}}
\def \oneTo#1{{[1..#1]}}
\def \epsPerm{{\epsilon_{\infty}}}
\def \epsOne{{\epsilon_1}}
\def \fori{{\mathbf{f}}}
\def \fest{{\mathbf{\hat{f}}}}
\def \fobs{{\mathbf{\tilde{f}}}}
\def \calM{{\mathcal{M}}}
\def \calD{{\mathcal{D}}}
\def \mse{{\mathrm{MSE}}}
\def \mae{{\mathrm{MAE}}}
\def \mfreqldpy{\texttt{multi-freq-ldpy}}

\begin{document}
\sloppy

\title{On the Utility Gain of Iterative Bayesian Update for Locally Differentially Private Mechanisms\thanks{Version of Record (DBSec'23): \url{https://doi.org/10.1007/978-3-031-37586-6_11}.}}
\titlerunning{On the Utility Gain of Iterative Bayesian Update for LDP Mechanisms}

\author{
Héber H. Arcolezi \and 
Selene Cerna \and 
Catuscia Palamidessi 
}

\authorrunning{H.H. Arcolezi, S. Cerna, \& C. Palamidessi
}

\institute{Inria and École Polytechnique (IPP), Palaiseau, France\\
\email{\{heber.hwang-arcolezi,selene-leya.cerna-nahuis,catuscia.palamidessi\}@inria.fr}
}
\maketitle              
\begin{abstract}
This paper investigates the utility gain of using Iterative Bayesian Update (IBU) for private discrete distribution estimation using data obfuscated with Locally Differentially Private (LDP) mechanisms. 
We compare the performance of IBU to Matrix Inversion (MI), a standard estimation technique, for seven LDP mechanisms designed for one-time data collection and for other seven LDP mechanisms designed for multiple data collections (\eg, RAPPOR). 
To broaden the scope of our study, we also varied the utility metric, the number of users $n$, the domain size $k$, and the privacy parameter $\epsilon$, using both synthetic and real-world data.
Our results suggest that IBU can be a useful post-processing tool for improving the utility of LDP mechanisms in different scenarios without any additional privacy cost. 
For instance, our experiments show that IBU can provide better utility than MI, especially in high privacy regimes (\ie, when $\epsilon$ is small). 
Our paper provides insights for practitioners to use IBU in conjunction with existing LDP mechanisms for more accurate and privacy-preserving data analysis. 
Finally, we implemented IBU for all fourteen LDP mechanisms into the state-of-the-art \mfreqldpy\ Python package (\url{https://pypi.org/project/multi-freq-ldpy/}) and open-sourced all our code used for the experiments as tutorials.

\keywords{Expectation-Maximization \and Iterative Bayesian Update \and Local Differential Privacy \and Distribution Estimation.}
\end{abstract}

\input{ms_text.tex}

\subsubsection{Acknowledgements} 
This work was supported by the European Research Council (ERC) project HYPATIA under the European Union’s Horizon 2020 research and innovation programme. Grant agreement n. 835294. 

%
%
%
\bibliographystyle{splncs04}
\bibliography{ms_bibliography.bib}
\end{document}

%% file: ms_text.tex
\section{Introduction} \label{sec:introduction}
The widespread availability of Big Data has led to the development of new methods for extracting valuable insights from large datasets.
However, the increased capacity to collect and analyze data also raises significant concerns about privacy, particularly in cases where sensitive information about individuals is involved.
Thus, the direct collection and storage of users' raw data on a centralized server should be avoided as these data are subject to illegal access~\cite{data_breaches} or internal fraud~\cite{cambridge}, for example. 
To address this issue, recent works have proposed several mechanisms satisfying Differential Privacy (DP)~\cite{Dwork2006,dwork2014algorithmic}, in the distributed setting, referred to as Local DP (LDP)~\cite{first_ldp}.

One of the strengths of LDP is its simple trust model: since each user sanitizes their data locally, user privacy is protected even if the server is malicious. 
However, the increased privacy of the local DP model comes at the cost of reduced utility.
That is, as each user's data is obfuscated, several lower bounds exist on the error of these LDP mechanisms. 
This strongly differentiates the local DP model from the central DP model.~\cite{cheu2021differential}.
To address this issue, one line of research has largely explored the issue of improving the utility of LDP distribution estimation mechanisms~\cite{tianhao2017,Bassily2015,wang2016mutual,Min2018,kairouz2016discrete,rappor}, in which the data collector estimates the number of users for each possible value based on the sanitized data of the users.
Distribution estimation is a primary objective of LDP~\cite{Xiong2020}, as it is a building block for more complex tasks (\eg, heavy hitter estimation~\cite{Bassily2015}, marginal estimation~\cite{Fanti2016,Liu2023,Ren2018,Filho2023}, and distribution monitoring~\cite{rappor,Arcolezi2023,Arcolezi2022_allomfree}).

Two commonly used estimators for distribution estimation under LDP are~\cite{Xiong2020}: Matrix Inversion (MI), which is based on simple linear algebra techniques, and Iterative Bayesian Update (IBU)~\cite{Agrawal2001,Agrawal2005}, which is based on the well-known Expectation-Maximization~\cite{Dempster1977} algorithm.
Thus, another line of research has explored how to improve the estimation at the server side through post-processing techniques~\cite{Wang2020_post_process,elsalamouny2022,ElSalamouny2020,Murakami2018}.
For instance, to improve the utility of the MI estimator, in~\cite{kairouz2016discrete}, the authors evaluate two post-processing techniques, namely, normalization (also adopted in this paper) and projection on the simplex.
More recently, Wang \etal~\cite{Wang2020_post_process} revisited and introduced, in total, ten post-processing techniques for the MI estimator.
Regarding IBU, ElSalamouny \& Palamidessi~\cite{ElSalamouny2020,elsalamouny2022} investigated a generalization of IBU for personalized LDP and, in~\cite{Murakami2018}, the authors improved IBU considering small number of users.

\textbf{Main contributions.}
While the aforementioned works have explored how to improve the estimation of LDP mechanisms on the server side through post-processing techniques, no study has compared the performance of MI and IBU estimators for more than three LDP mechanisms.
This paper aims to fill this research gap by conducting an in-depth empirical analysis of the utility gain by using IBU instead of MI for \textit{fourteen state-of-the-art LDP mechanisms}.
More precisely, we have experimentally assessed seven state-of-the-art LDP distribution estimation mechanisms, namely, Generalized Randomized Response (GRR)~\cite{kairouz2016discrete}, Binary Local Hashing (BLH)~\cite{Bassily2015}, Optimal Local Hashing (OLH)~\cite{tianhao2017}, Symmetric Unary Encoding (SUE)~\cite{rappor}, Optimal Unary Encoding (OUE)~\cite{tianhao2017}, Subset Selection (SS)~\cite{wang2016mutual,Min2018}, and Thresholding with Histogram Encoding (THE)~\cite{tianhao2017}. 
In addition, we also extended the analysis to seven state-of-the-art memoization-based LDP mechanisms, namely, Longitudinal GRR (L-GRR)~\cite{Arcolezi2022_allomfree}, four Longitudinal Unary Encoding (L-UE)~\cite{rappor,Arcolezi2022_allomfree} mechanisms, and two Longitudinal Local Hashing (L-LH)~\cite{Arcolezi2023} mechanisms.
To further broaden the scope of our study, we have evaluated two popular utility metrics by varying the number of users $n$, the privacy guarantee $\epsilon$, and the domain size $k$, with both synthetic and real-world datasets.
Finally, this paper open-sources the implementation of IBU for all aforementioned LDP mechanisms into the \mfreqldpy\ Python package~\cite{Arcolezi2022_multi_freq_ldpy}, as well as the code used for our experiments in the form of tutorials.

In summary, the three main contributions of this paper are:

\begin{itemize}
    \item We present an in-depth analysis of the utility gain of IBU over MI for seven state-of-the-art LDP mechanisms designed for one-time data collection;
    
    \item To the authors' knowledge, we are the first to extend IBU and conduct a detailed analysis of its utility gain over MI for seven state-of-the-art longitudinal LDP mechanisms designed for multiple data collections;
    
    \item We open-sourced the IBU implementation for all aforementioned LDP mechanisms into \mfreqldpy~\cite{Arcolezi2022_multi_freq_ldpy}, facilitating future research in the area and making it easier for researchers to replicate and build upon our findings.
\end{itemize}

\noindent \textbf{Outline.} The rest of this paper is organized as follows.
Section~\ref{sec:preliminaries} introduces the notation, the problem, and the metrics, and briefly reviews LDP and the two estimators, \ie, MI and IBU.
In Section~\ref{sec:ldp_mechanisms}, we present all LDP distribution estimation mechanisms analyzed in this paper.
Next, Section~\ref{sec:results_discussion} details the experimental setting and main results before discussing related work in Section~\ref{sec:rel_work}.
Finally, we conclude this work indicating future perspectives in Section~\ref{sec:conclusion}.

\section{Preliminaries} \label{sec:preliminaries}
In this section, we present the notation, the problem, the utility metrics, the LDP privacy model, and both MI and IBU estimators, used in this paper.

\subsection{Notations \& Problem Statement} \label{sub:problem_statement}

The main notation used throughout this paper is summarized in Table~\ref{tab:notation}.
This paper considers a distributed setting with $n$ users and one untrusted server. 
Each user holds a value $v \in V$, where $V$ ranges on a finite domain $\calD$ of size $k$. 
The distribution of $V$ is represented by $\fori=\{f(v)\}_{v\in \calD} \in [0,1]^k$, \ie, $\fori(v)$ is the probability of $v \in \calD$. 
To satisfy LDP, each user $i$, for $i \in \oneTo{n}$, will apply an obfuscation mechanism $\calM_{(\epsilon)}$ to their value $v^i$ to obtain an output $y^i=\calM_{(\epsilon)}\left(v^i\right)$ from a finite set $Y$ ranging on $\calD$.
The obfuscation mechanism $\calM_{(\epsilon)}$ maps $v \in \calD$ to $y \in \calD$ through a channel matrix $A_{vy}$, \ie, the probability that $\calM_{(\epsilon)}$ yields $y$ from $v$.
Upon collecting the obfuscated data of all users, the server computes the observed distribution of $y^i$ denoted as $\fobs=\{\tilde{f}(v)\}_{v\in \calD} \in [0,1]^k$, where $\fobs(v)$ is the probability of $v \in \calD$ and is calculated by counting the number of times that $v$ is observed in $Y$ and divided by $n$.
The server's goal is to estimate $\fori$ by calculating $\fobs$ and with the knowledge of the obfuscation mechanism $\calM_{(\epsilon)}$.
We denote the estimated distribution of $\fori$ by $\fest=\{\hat{f}(v)\}_{v\in \calD}$. 

\setlength{\tabcolsep}{6pt}
\begin{table}[ht]
    \centering
    \begin{tabular}{c l}
    \toprule
     Symbol                 & Description \\
     \toprule
     $\calD$                & Data domain. \\
     $k$                    & Domain size $k=|D|$. \\
     $n$                    & Number of users. \\
     $V$                    & Discrete set of original data $\{v^1, v^2, v^3, \ldots, v^n\}$, ranging on $\calD$. \\
     $Y$                    & Discrete set of obfuscated data $\{y^1, y^2, y^3, \ldots, y^n\}$, ranging on $\calD$. \\
     $\fori$                & Distribution of original data, where $\fori=\{f(v)\}_{v\in \calD} \in [0,1]^k$. \\
     $\fobs$                & Distribution of obfuscated data, where $\fobs=\{\tilde{f}(v)\}_{v\in \calD} \in [0,1]^k$. \\
     $\fest$                & Estimated distribution of $\fori$, where $\fest=\{\hat{f}(v)\}_{v\in \calD}$. \\ 
     $\calM_{(\epsilon)}$   & $\epsilon$-LDP obfuscation mechanism, $\calM_{(\epsilon)} : V \to Y$. \\
     $A_{vy}$               & Channel matrix, probability that $\calM_{(\epsilon)}$ yields $y \in \calD$ from $v \in \calD$. \\
     $\Gamma$               & Utility gain of IBU over MI defined in Eq.~\eqref{eq:utility_gain}. \\ 
     $[a, b]$               & Set of all real numbers $\geq a$ and $\leq b$. \\
     $\oneTo{a}$            & Set of integers $\{1, 2, 3, \ldots,a\}$. \\     
     $\mathbf{a}_i$         & $i$-th coordinate of vector $\mathbf{a}$. \\ 
     \bottomrule
    \end{tabular}
    \caption{Notations}
    \label{tab:notation}
\end{table}

As utility metrics, we use the Mean Squared Error (MSE) and the Mean Absolute Error (MAE) to measure the difference between the original distribution $\fori$ and estimated distribution $\fest$, which are formally defined as:

\begin{itemize}
    \item \textbf{MSE.} Given $\fori$ and $\fest$, the MSE is an expectation of the squared error:

    \begin{equation*}
        \mse = \frac{1}{k} \sum_{v\in \calD} (f(v) - \hat{f}(v))^2 \textrm{.}
    \end{equation*}

    \item \textbf{MAE.} Given $\fori$ and $\fest$, the MAE is an expectation of the absolute error:

    \begin{equation*}
        \mae = \frac{1}{k} \sum_{v\in \calD} | f(v) - \hat{f}(v) | \textrm{.}
    \end{equation*}
\end{itemize}

In this paper, we are interested in comparing the two main statistical estimators for $\fest$, namely, MI and IBU.
Based on the two popular metrics MSE and MAE, we define the utility gain $\Gamma$ of IBU over MI as follows:

\begin{equation} \label{eq:utility_gain}
    \Gamma (\%) = 100 \cdot \max \left( \frac{\mathrm{Metric}_{\mathrm{MI}} - \mathrm{Metric}_{\mathrm{IBU}}}{\mathrm{Metric}_{\mathrm{MI}}} , 0\right) \textrm{,}
\end{equation}

\noindent that is, the utility gain is clipped to zero to represent no gain.

\subsection{Local Differential Privacy (LDP)} \label{sub:ldp}

In this paper, we use LDP~\cite{first_ldp} as the privacy model considered, formalized as:

\begin{definition}[$\epsilon$-Local Differential Privacy]\label{def:ldp} A randomized mechanism ${\calM}$ satisfies $\epsilon$-LDP, where $\epsilon>0$, if for any input $v, v' \in \calD$, we have:

\begin{equation*} \label{eq:ldp}
    \forall y \in \calD : \quad    \Pr[{\calM}(v) = y] \leq e^\epsilon \cdot \Pr[{\calM}(v') = y]  \textrm{.}
\end{equation*}

\end{definition}

\begin{proposition}[Post-Processing~\cite{dwork2014algorithmic}]
\label{prop:post-processing}
If $\calM$ is $\epsilon$-LDP, then for any function $g$, the composition of $\calM$ and $g$, \ie, $g (\calM)$ satisfies $\epsilon$-LDP.
\end{proposition}

The parameter $\epsilon$ controls the privacy-utility trade-off.
Note that lower values of $\epsilon$ result in tighter privacy protection and vice versa. 
Moreover, an LDP mechanism is called ``pure LDP" if it satisfies the following definition~\cite{tianhao2017}.

\begin{definition}[Pure LDP~\cite{tianhao2017}]\label{def:pure_ldp} An $\epsilon$-LDP mechanism is pure if there are two probability parameters $0 < q^* < p^* < 1$ such that for all $v \in \calD$,

\begin{equation*}
    \begin{aligned}
                                & \quad \Pr[\calM(v) \in \{y | v \in S(y)\}] = p^*,\\
        \forall{v \neq v' \in \calD}      & \quad \Pr[\calM(v') \in \{y | v \in S(y)\}] = q^*,
    \end{aligned}
\end{equation*}

\noindent where $S(y)$ is the set of items that $y$ supports.
That is, $S(y)$ maps each possible output $y$ to a set of input values that $y$ ``supports".

\end{definition}

\subsection{Matrix Inversion (MI) Estimator} \label{sub:mi}

All obfuscation mechanisms we analyze in this paper are pure LDP (\cf{} Definition~\ref{def:pure_ldp}), which makes their analysis advantageous.
For instance, for any pure LDP mechanism, the server estimates the distribution $\fest=\{\hat{f}(v)\}_{v\in \calD} \in \mathbb{R}^k$ as~\cite{tianhao2017}:

\begin{equation} \label{eq:est_matrix_inversion}
    \forall{v\in \calD}, \quad \hat{f}(v) = \frac{\sum_{i=1}^n \mathds{1}_{S(y^i)}(v^i) - nq^*}{n(p^* - q^*)} \textrm{,}
\end{equation}

\noindent in which $y^i$ represents the obfuscated data provided by each user (for $i \in \oneTo{n}$), and the function $\mathds{1}_{S(y^i)}(v^i)$ equals $1$ if $y^i$ supports the value $v^i$, and $0$ otherwise.
However, when the number of users is small, many elements estimated with Eq.~\eqref{eq:est_matrix_inversion} can be negative. 
Therefore, we clip the negative estimates to $0$ and re-normalize the estimated frequencies so that they sum up to $1$, ensuring that they constitute a valid probability distribution.
Another way to write Eq.~\eqref{eq:est_matrix_inversion} is: $\fest = \fobs A_{vy}^{-1}$, in which $\fobs$ is the observed distribution and $A_{vy}^{-1}$ is the inverse of $A_{vy}$~\cite{ElSalamouny2020,kairouz2016discrete,Murakami2018}.
For any pure LDP mechanism, we observed that the channel matrix $A_{vy}$ is a square matrix with values $p^*$ in the diagonal and $q^*$ otherwise.

\subsection{Iterative Bayesian Update (IBU) Estimator} \label{sub:ibu}

The IBU estimator~\cite{Agrawal2001,Agrawal2005,ElSalamouny2020} is based on the Expectation Maximization (EM)~\cite{Dempster1977} method, which is a powerful method for obtaining parameter estimates when part of the data is missing. 
IBU estimates the distribution of $\fori$ as follows. 
Let $\fobs$ be the observed distribution on $Y$, $A_{vy}$ be the channel matrix, which represents the probability of obtaining $y$ from $v$, and $\fest$ be the estimated distribution on $V$. 
IBU starts with a full-support distribution $\fest^0$ on $V$ (\eg, the uniform distribution). Next, it iteratively generates new distributions by updating Eq.~\eqref{eq:est_ibu}, where $(*)$, $(\cdot)$, and $(/)$ represent the element-wise product, dot product, and element-wise division, respectively. 
Finally, it converges until a specified tolerance is reached.

\begin{equation} \label{eq:est_ibu} 
    \fest^{t+1} = {\fobs} \cdot \frac{{\fest^t} * A_{vy}}{{\fest^t} \cdot A_{vy}}  \mathrm{.}
\end{equation}

\section{LDP Distribution Estimation Mechanisms} \label{sec:ldp_mechanisms}

In this section, we briefly review fourteen state-of-the-art LDP distribution estimation mechanisms, \ie, seven for one-time collection (Section~\ref{sub:one_time_ldp_mechanisms}) and seven for multiple collections (Section~\ref{sub:long_ldp_mechanisms}). 

\subsection{LDP Mechanisms for One-Time Data Collection} \label{sub:one_time_ldp_mechanisms}
Algorithm~\ref{alg:general_LDP} exhibits the general procedure for one-time discrete distribution estimation under pure LDP.
The five following subsections briefly present state-of-the-art pure LDP mechanisms.

\begin{algorithm}
\centering
\caption{General pure LDP procedure for distribution estimation.}
\label{alg:general_LDP}
\begin{algorithmic}[1]
\Statex \textbf{Input :} Original data of users, privacy parameter $\epsilon$, mechanism $\calM_{(\epsilon)}$.
\Statex \textbf{Output :} Estimated discrete distribution.

\Statex \textcolor{blue}{\# \texttt{User-side}}

\State \textbf{for} each user $i \in \oneTo{n}$ with input data $v^{i} \in V$ \textbf{do}

\State  \hskip1em \textcolor{magenta}{\texttt{Encode}}($v^{i}$) into a specific format (\textbf{if needed});
\State  \hskip1em \textcolor{magenta}{\texttt{Obfuscate}}($v^{i}$) as $y^{i}=\calM_{(\epsilon)}(v^{i})$;
\State  \hskip1em Transmit $y^{i}$ to the aggregator. 

\State \textbf{end for}

\Statex \textcolor{blue}{\# \texttt{Server-side}}
\State Obtain the support set $S(y)$ and probabilities $p^*$ and $q^*$ for $\calM_{(\epsilon)}$. 
\State \textcolor{magenta}{\texttt{Estimate}} Aggregate the obfuscated data $y^{i}$ ($i\in\oneTo{n}$) to estimate $\{\hat{f}(v)\}_{v\in \calD}$.
\State \textbf{return :} Estimated discrete distribution $\fest$ (\ie, a $k$-bins histogram).
\end{algorithmic}
\end{algorithm}

\subsubsection{Generalized Randomized Response (GRR)} 
\label{sub:GRR}

The GRR~\cite{kairouz2016discrete} mechanism generalizes the Randomized Response (RR) surveying technique proposed by Warner~\cite{Warner1965} for $k \geq 2$ while satisfying $\epsilon$-LDP.

\noindent \textbf{Encode.} GRR~\cite{kairouz2016discrete} uses no particular encoding, \ie, $\texttt{Encode}_{\mathrm{GRR}}(v)=v$.

\noindent \textbf{Obfuscate.} Given the user's personal data $v$, GRR outputs $v$ with probability $p$, and any other randomly chosen value $v' \in \calD \setminus \{v\}$ otherwise. 
More formally:

\begin{equation} \label{eq:grr}
    \forall{y \in \calD}  : \quad \Pr[\calM_{\mathrm{GRR}(\epsilon, k)}(v)=y] = \begin{cases} p=\frac{e^{\epsilon}}{e^{\epsilon}+k-1} , \textrm{ if } y = v, \\ q=\frac{1}{e^{\epsilon}+k-1}, \textrm{ otherwise} \textrm{,} \end{cases}
\end{equation}

\noindent in which $y$ is the obfuscated value sent to the server. 

\noindent \textbf{Estimate.} For aggregation at the server side, it is important to obtain the support set $S(y)$ and the probabilities $p^*$ and $q^*$. 
For GRR, an obfuscated value $y$ only supports itself, \ie, $S_{\mathrm{GRR}}(y)=\{y\}$.
Given the support set and with $p^*=p$ and $q^*=q$, the server can estimate item frequencies using Eq.~\eqref{eq:est_matrix_inversion} for MI and Eq.~\eqref{eq:est_ibu} for IBU.

\subsubsection{Local Hashing (LH)} \label{sub:_mechanisms}

LH mechanisms~\cite{tianhao2017,Bassily2015} can handle a large domain size $k$ by using hash functions to map an input data to a smaller domain of size $g$, for $2 \leq g \ll k$, and then applying GRR to the hashed value. 
Let $\mathscr{H}$ be a universal hash function family such that each hash function $\mathrm{H} \in \mathscr{H}$ hashes a value $v \in \calD$ into $\oneTo{g}$, \ie, $\mathrm{H} : \calD \rightarrow \oneTo{g}$. 
There are two variations of LH mechanisms:

\begin{itemize}
    \item \textbf{Binary LH (BLH)~\cite{Bassily2015}.} A simple case that just sets $g=2$;
    \item \textbf{Optimal LH (OLH)~\cite{tianhao2017}.} An optimized version that selects $g=\lfloor e^{\epsilon} + 1 \rceil$.
\end{itemize}

\noindent \textbf{Encode.} $\texttt{Encode}_{\textrm{LH}}(v)=\langle \mathrm{H}, \mathrm{H}(v) \rangle$, in which $\mathrm{H} \in \mathscr{H}$ is selected at random.

\noindent \textbf{Obfuscate.} LH mechanisms only obfuscate the hash value $\mathrm{H}(v)$ with GRR and does not change $\mathrm{H}$. 
In particular, the LH reporting mechanism is:

\begin{equation*}
    \calM_{\mathrm{LH}(\epsilon)}(v) \coloneqq \langle \mathrm{H}, \calM_{\mathrm{GRR}(\epsilon, g)}(\mathrm{H}(v)) \rangle \textrm{,}
\end{equation*}

\noindent in which $\calM_{\mathrm{GRR}(\epsilon, g)}$ is given in Eq.~\eqref{eq:grr}, while operating on the new domain $\oneTo{g}$. 
Each user reports, the hash function and obfuscated value $\langle \mathrm{H}, y \rangle$ to the server. 

\noindent \textbf{Estimate.} The support set for LH mechanisms is $S_{\mathrm{LH}}\left( \langle \mathrm{H}, y \rangle \right) = \{v | \mathrm{H}(v) = y \}$, \ie, the set of values that are hashed into the obfuscated data $y$.
LH mechanisms are pure with probabilities $p^* = p$ and $q^* = \frac{1}{g}$~\cite{tianhao2017}.
Thus, the server can estimate item frequencies using Eq.~\eqref{eq:est_matrix_inversion} for MI and Eq.~\eqref{eq:est_ibu} for IBU.

\subsubsection{Unary Encoding (UE)} \label{sub:ue_mechanisms}

UE mechanisms~\cite{rappor,tianhao2017} interpret the user's input data $v \in \calD$, as a one-hot $k$-dimensional vector and obfuscate each bit independently.

\noindent \textbf{Encode.} $\texttt{Encode}_{\mathrm{UE}}(v)=\textbf{v}=[0, \ldots, 0, 1, 0, \ldots, 0]$ is a binary vector with only the bit at the position corresponding to $v$ set to $1$ and the other bits set to $0$. 

\noindent \textbf{Obfuscate.} The obfuscation function of UE mechanisms randomizes the bits from $\textbf{v}$ independently to generate $\textbf{y}$ as follows:

\begin{equation}  \label{eq:ue_parameters}
    \forall{i \in \oneTo{k}} : \quad \Pr[\textbf{y}_i=1] =\begin{cases} p, \textrm{ if } \textbf{v}_i=1 \textrm{,} \\ q, \textrm{ if } \textbf{v}_i=0 \textrm{,}\end{cases}
\end{equation}

\noindent in which $\textbf{y}$ is sent to the server. 
There are two variations of UE mechanisms:

\begin{itemize}
    \item \textbf{Symmetric UE (SUE)~\cite{rappor}.} Also known as Basic One-Time RAPPOR, SUE selects $p=\frac{e^{\epsilon/2}}{e^{\epsilon/2}+1}$ and $q=\frac{1}{e^{\epsilon/2}+1}$ in Eq.~\eqref{eq:ue_parameters}, such that $p+q=1$;
    \item \textbf{Optimal UE (OUE)~\cite{tianhao2017}.} OUE selects $p=\frac{1}{2}$ and $q=\frac{1}{e^{\epsilon}+1}$ in Eq.~\eqref{eq:ue_parameters}.
\end{itemize}

\noindent \textbf{Estimate.} An obfuscated vector $\textbf{y}$ supports an item $v$ if and only if the $v$-th bit of $\textbf{y}$, denoted as $\textbf{y}_v$, equals to 1. 
Formally, we have $S_{\mathrm{UE}}(\textbf{y})=\{i | \textbf{y}_i = 1\}$, for $i \in \oneTo{k}$.
UE mechanisms are pure with $p^*=p$ and $q^*=q$.
Thus, the server can estimate item frequencies using Eq.~\eqref{eq:est_matrix_inversion} for MI and Eq.~\eqref{eq:est_ibu} for IBU.

\subsubsection{Subset Selection (SS)} \label{sub:SS}

The SS~\cite{wang2016mutual,Min2018} mechanism was proposed for the case where the obfuscation output is a subset of values $\Omega$ of the original domain $V$.
The user's true value $v \in \calD$ has higher probability of being included in the subset $\Omega$, compared to the other values in $\calD \setminus \{v\}$. 
The optimal subset size that minimizes the MSE is $\omega= \lfloor \frac{k}{e^{\epsilon}+1} \rceil$. 

\noindent \textbf{Encode.} $\texttt{Encode}_{\mathrm{SS}}(v)=\Omega=\emptyset$ is an empty subset. 

\noindent \textbf{Obfuscate.} Given the empty subset $\Omega$, the true value $v$ is added to $\Omega$ with probability $p=\frac{\omega e^{\epsilon}}{\omega e^{\epsilon} + k - \omega}$. 
Finally, it adds values to $\Omega$ as follows:

\begin{itemize}
    \item If $v \in \Omega$, then $\omega - 1$ values are sampled from $\calD \setminus \{v\}$ uniformly at random (without replacement) and are added to $\Omega$;
    
    \item If $v \notin \Omega$, then $\omega$ values are sampled from $\calD \setminus \{v\}$ uniformly at random (without replacement) and are added to $\Omega$.
\end{itemize}

Afterward, the user sends the subset $\Omega$ to the server.

\noindent \textbf{Estimate.} An obfuscated subset $\Omega$ supports an item $v$ if and only if $v \in \Omega$. 
Formally, we have $S_{\mathrm{SS}}(\Omega)=\{v | v \in \Omega\}$.
The SS mechanism is pure with $p^*=p$ and $q^*= \frac{\omega e^{\epsilon}(\omega-1) + (k - \omega)\omega}{(k - 1)(\omega e^{\epsilon} + k - \omega)}$.
Thus, the server can estimate item frequencies using Eq.~\eqref{eq:est_matrix_inversion} for MI and Eq.~\eqref{eq:est_ibu} for IBU.

\subsubsection{Thresholding with Histogram Encoding (THE)} \label{sub:the}

In Histogram Encoding (HE)~\cite{tianhao2017}, an input data $v \in \calD$ is encoded as a one-hot $k$-dimensional histogram and each bit is obfuscated independently with the Laplace mechanism~\cite{Dwork2006}.
For any two input data $v,v' \in \calD$, the L1 distance between the two vectors is $\Delta=2$. 

\noindent \textbf{Encode.} $\texttt{Encode}_{\mathrm{HE}(v)} = \textbf{v} = [0.0, \ldots, 0.0 , 1.0, 0.0, \ldots , 0.0]$, where only the $v$-th component is $1.0$ and the other bits are set to $0.0$.

\noindent \textbf{Obfuscate.} HE outputs $\textbf{y}$ such that $\textbf{y}_i = \textbf{v}_i + \textrm{Lap}\left( \frac{2}{\epsilon} \right)$, in which $\mathrm{Lap}(\beta)$ is the Laplace distribution where $\Pr[\mathrm{Lap}\left( \beta \right)=x]=\frac{1}{2\beta}e^{-|x|/\beta}$.

\noindent \textbf{Estimate.} Since Laplace noise with zero mean is added to each bit independently, a simple aggregation method is to sum the noisy counts for each bit.
However, this aggregation method does not provide a support function and is not pure (a.k.a. SHE in~\cite{tianhao2017}).
Wang \etal~\cite{tianhao2017} proposed THE such that the user reports (or the server computes) the support set as $S_{\mathrm{THE}}(\textbf{y}) = \{ v \hspace{0.1cm} | \hspace{0.1cm} \textbf{y}_v \hspace{0.1cm} >  \hspace{0.1cm}\theta\}$, \ie, each noise count that is $> \theta$ supports the corresponding value.
According to~\cite{tianhao2017}, the optimal threshold value $\theta$ that minimizes the $\mse=\frac{2 e^{\epsilon \theta / 2} - 1}{(1 + e^{\epsilon(\theta - 1/2)} - 2e^{\epsilon \theta / 2})^2}$ is within $(0.5, 1)$.
THE is pure with $p^*=1 - \frac{1}{2}e^{\frac{\epsilon}{2}(\theta - 1)}$ and $q^*= \frac{1}{2}e^{-\frac{\epsilon}{2}\theta}$.
Thus, the server can estimate item frequencies using Eq.~\eqref{eq:est_matrix_inversion} for MI and Eq.~\eqref{eq:est_ibu} for IBU.

\subsection{LDP Mechanisms for Multiple Data Collections} \label{sub:long_ldp_mechanisms}

The LDP mechanisms presented in Section~\ref{sub:one_time_ldp_mechanisms} consider a one-time data collection. 
In this section, we consider the server is interested in multiple collections throughout time. 
More precisely, for longitudinal data (\ie, data that is monitored over time $t \in [\tau]$), all major LDP mechanisms make use of an internal memoization mechanism~\cite{rappor}. 
Memoization was designed to enable longitudinal collections through memorizing an obfuscated version of the true value $v$ as $y=\calM_{\epsPerm}(v)$, and consistently reusing $y$ as the input to a second step of obfuscation in time $t \in [\tau]$. 
Algorithm~\ref{alg:general_LDP_long} exhibits the general procedure for longitudinal discrete distribution estimation under LDP guarantees.

Without loss of generality, we consider $\tau=1$ (one-time collection) in this paper, which will allow us to compare all LDP mechanisms under the same privacy guarantees: (i) $\epsPerm$-LDP, which is the privacy guarantees of the first obfuscation step and the upper bound when $\tau \rightarrow \infty$; (ii) $\epsOne$-LDP, which is the privacy guarantees of the first report by chaining two LDP mechanisms (\ie, the lower bound when $\tau=1$).
Naturally, $\epsOne \leq \epsPerm$ and, in practice, we want $\epsOne \ll \epsPerm$.

The three following subsections briefly present state-of-the-art longitudinal LDP mechanisms~\cite{rappor,Arcolezi2022_allomfree,Arcolezi2023}, which are based on GRR, LH, and UE mechanisms, \ie, same encoding, perturbation, and aggregation methods.
Therefore, we focus on presenting the composition of two LDP mechanisms (\ie, $\calM_1 \circ \calM_2$) and both $p^*$ and $q^*$ parameters that are based on the privacy guarantees $\epsPerm,\epsOne$.

\begin{algorithm}
\centering
\caption{Memoization-based procedure for longitudinal distribution estimation under LDP guarantees~\cite{rappor,Arcolezi2022_allomfree,Arcolezi2023}.}
\label{alg:general_LDP_long}
\begin{algorithmic}[1]
\Statex \textbf{Input :} Original data of users, privacy parameters $\epsPerm,\epsOne$, mechanisms $\calM_1$, $\calM_2$.
\Statex \textbf{Output :} Estimated discrete distribution $\fest$ at each $t \in [\tau]$.

\Statex \textcolor{blue}{\# \texttt{User-side}}

\State \textbf{for} each user $i \in \oneTo{n}$ with input data $v^{i} \in V$ \textbf{do}

\State  \hskip1em \textcolor{magenta}{\texttt{Encode}}($v^{i}$) into a specific format (\textbf{if needed});
\State  \hskip1em \textcolor{magenta}{\texttt{Obfuscate}}($v^{i}$) as $y^{i}=\calM_{1(\epsPerm)}(v^{i})$; \Comment{First obfuscation step: $p_1^*$ and $q_1^*$}
\State  \hskip1em \textcolor{brown}{\texttt{Memoize}}($y^{i}$) for $v^{i}$.
\State  \hskip1em \textbf{for} each time $t \in [\tau]$ \textbf{do}:
\State  \hskip3em \textcolor{magenta}{\texttt{Obfuscate}}($y^{i}$) as $y_t^{i}=\calM_{2(\epsilon)}(y^{i})$; \Comment{Second obfuscation step: $p_2^*$ and $q_2^*$}
\State  \hskip3em Transmit $y_t^{i}$ to the aggregator. 
\State  \hskip1em \textbf{end for}

\State \textbf{end for}

\Statex \textcolor{blue}{\# \texttt{Server-side}}
\State Obtain the support set $S(y)$ and probabilities $p_1^*$, $q_1^*$, $p_2^*$, and $q_2^*$ for $\calM_{1(\epsilon)}$, $\calM_{2(\epsilon)}$. 
\State \textbf{for} each time $t \in [\tau]$ \textbf{do}:
\State \hskip1em \textcolor{magenta}{\texttt{Estimate}} Aggregate the obfuscated data $y_t^{i}$ ($i\in\oneTo{n}$) to estimate $\{\hat{f}(v)\}_{v\in \calD}$.
\State \textbf{end for}
\end{algorithmic}
\end{algorithm}

\subsubsection{Longitudinal GRR (L-GRR)} \label{sub:l_grr}

The L-GRR~\cite{Arcolezi2022_allomfree} mechanism chains GRR in both obfuscation steps of Algorithm~\ref{alg:general_LDP_long} (\ie, GRR $\circ$ GRR).

\textbf{L-GRR Parameters:}
\begin{itemize}
    \item First obfuscation step: $p_1^* = \frac{e^{\epsPerm}}{e^{\epsPerm}+k-1}$ and $q_1^*=\frac{1-p_1}{k-1}$;

    \item Second obfuscation step: $p_2^* = \frac{e^{\epsPerm + \epsOne} - 1}{- k e^\epsOne + \left(k - 1\right) e^{\epsPerm} + e^\epsOne + e^{\epsOne + \epsPerm} - 1}$ and $q_2^*=\frac{1-p_2}{k-1}$.

    \item Aggregation: $p^* = p_1^*  p_2^* + q_1^*  q_2^*$ and $q^* = p_1^*  q_2^* + q_1^*  p_2^*$.
\end{itemize}

\subsubsection{Longitudinal UE (L-UE) mechanisms} \label{sub:l_ue} 

Arcolezi \etal~\cite{Arcolezi2022_allomfree} analyzed all four combinations between OUE and SUE in both obfuscation steps, \ie, L-SUE (SUE $\circ$ SUE, a.k.a. Basic RAPPOR~\cite{rappor}), L-SOUE (SUE $\circ$ OUE), L-OUE (OUE $\circ$ OUE), and L-OSUE (OUE $\circ$ SUE).
Without loss of generality, we present the parameters of L-OSUE only and refer the readers to~\cite{Arcolezi2022_allomfree} and to \mfreqldpy~\cite{Arcolezi2022_multi_freq_ldpy} to access the parameters of the other L-UE mechanisms.

\textbf{L-OSUE Parameters:}
\begin{itemize}
    \item First obfuscation step: $p_1^* = \frac{1}{2}$ and $q_1^*=\frac{1}{e^{\epsPerm}+1}$;

    \item Second obfuscation step: $p_2^* = \frac{e^\epsPerm e^\epsOne - 1}{e^\epsPerm - e^\epsOne + e^{\epsPerm + \epsOne} - 1}$ and $q_2^*=1 - p_2$.

    \item Aggregation: $p^* = p_1^* p_2^* + (1 - p_1^*) q_2^*$ and $q^* = q_1^* p_2^* + (1 - q_1^*) q_2^*$.
\end{itemize}

\subsubsection{Longitudinal LH (L-LH)} \label{sub:l_lh} 

Arcolezi \etal~\cite{Arcolezi2023} extended LH mechanisms for two obfuscation steps.
More specifically, L-LH mechanisms use a hash function $\mathrm{H} \in \mathscr{H}$ to map a value $v \in \calD \to \oneTo{g}$ and use L-GRR to obfuscate the hash value $\mathrm{H}(v)$ in the new domain $\oneTo{g}$.
There are two variations of L-LH:
\begin{itemize}
    \item L-BLH: A simple case that just sets $g=2$;
    \item L-OLH: Let $a=e^{\epsPerm}$ and $b=e^{\epsOne}$, this optimized version selects $g=1{+}{\max}\left(1,\left \lfloor \frac{1 {-} a^{2} {+} \sqrt{a^{4} {-} 14 a^{2} {+} 12 a b (1{-} a b){+} 12 a^3 b {+} 1} }{6 (a {-} b)} \right \rceil \right )$.
\end{itemize}

\textbf{L-LH Parameters:}
\begin{itemize}
    \item First obfuscation step: $p_1^* = \frac{e^{\epsPerm}}{e^{\epsPerm}+g-1}$ and $q_1^*=\frac{1}{g}$;

    \item Second obfuscation step: $p_2^* = \frac{e^{\epsPerm + \epsOne} - 1}{- g e^\epsOne + \left(g - 1\right) e^{\epsPerm} + e^\epsOne + e^{\epsOne + \epsPerm} - 1}$ and $q_2^*=\frac{1-p_2}{g-1}$.

    \item Aggregation: $p^* = p_1^*  p_2^* + q_1^*  q_2^*$ and $q^* = p_1^*  q_2^* + q_1^*  p_2^*$.
\end{itemize}

\section{Experimental Evaluation}  \label{sec:results_discussion}

In this section, we present the setting of our experiments and our main results.

\subsection{Setup of Experiments} \label{sub:setup_experiments}

\noindent \textbf{Environment.} All algorithms are implemented in Python 3 and run on a local machine with 2.50GHz Intel Core i9 and 64GB RAM. 
The codes we develop for all experiments are available in the \textit{tutorials} folder of the \mfreqldpy\ GitHub repository (\url{https://github.com/hharcolezi/multi-freq-ldpy})~\cite{Arcolezi2022_multi_freq_ldpy}. 

\noindent \textbf{IBU parameters.} We fix the \# iterations to $10000$ and the tolerance to $10^{-12}$.

\noindent \textbf{Data distribution.} For ease of reproducibility, we generate synthetic data following five well-known distributions and use one real-world dataset.

\begin{itemize}

    \item \textbf{Gaussian.} We generate $n$ samples following the Gaussian distribution with parameters $\mu=1000$ and $\sigma^2=100$ and bucketize to a $k$-bins histogram;
    
    \item \textbf{Exponential.} We generate $n$ samples following the Exponential distribution with parameters $\lambda=1$ and bucketize to a $k$-bins histogram;
    
    \item \textbf{Uniform.} We generate $n$ samples following the Uniform distribution in range $[100,10000]$ and bucketize to a $k$-bins histogram;
    
    \item \textbf{Poisson.} We generate $n$ samples following the Poisson distribution with parameters $\lambda=5$ and bucketize to a $k$-bins histogram;
    
    \item \textbf{Triangular.} We generate $n$ samples following the Triangular distribution with parameters $a=100$ (right), $c=4500$ (center), and $b=10000$ (left) and bucketize to a $k$-bins histogram;

    \item \textbf{Real.} We query for $n$ samples of the Income dataset, which is retrieved with the \texttt{folktables}~\cite{ding2021retiring} Python package.
    We only use the ``PINCP" numerical attribute and set the parameters: \texttt{survey\_year}=`2018', \texttt{horizon}=`5-Year', \texttt{survey}=`person'.
    Afterward, we bucketize to a $k$-bins histogram.
\end{itemize}

\noindent \textbf{Varying parameters.} When generating our data, we considered different:

\begin{itemize}
    \item \textbf{Domain size.} We varied the domain size $k$ of attributes as $k \in \{2, 50, 100, 200\}$;

    \item \textbf{Number of users.} We varied the number of users $n$ as $n \in \{20000, 100000\}$.

\end{itemize}

\noindent \textbf{Methods evaluated.} We assessed the following LDP mechanisms for:

\begin{itemize}
    \item \textbf{One-time collection.} All seven pure LDP mechanisms from Section~\ref{sub:one_time_ldp_mechanisms}, \ie, GRR, SUE, OUE, SS, THE, BLH, and OLH;

    \item \textbf{Multiple collections.} All seven longitudinal pure LDP mechanisms from Section~\ref{sub:long_ldp_mechanisms}, \ie, L-GRR, L-SUE, L-SOUE, L-OUE, L-OSUE, L-BLH, and L-OLH.

\end{itemize}

\noindent \textbf{Stability.} As LDP protocols and the synthetic data generation procedure are randomized, we report average results over 20 runs.

\noindent \textbf{Metrics.} We evaluated the privacy-utility trade-off of all fourteen LDP mechanisms according to:

\begin{itemize}
    \item \textbf{Utility metrics.} We measured the utility gain of IBU over MI following Eq.~\eqref{eq:utility_gain} for both MSE and MAE metrics;

    \item \textbf{Privacy guarantees.} We varied $\epsilon$ of LDP mechanisms for one-time collection as $\epsilon \in \{1, 2, 4\}$, representing high, medium, and low privacy regimes, respectively.
    Besides, we varied the privacy guarantees $\epsPerm$ and $\epsOne$ of longitudinal LDP mechanisms for multiple collections as $\epsPerm \in \{2, 4, 8\}$ and $\epsOne=\frac{\epsPerm}{2}$.
    That is, the first report has the same privacy guarantees as one-time LDP mechanisms, and, thus, similar results are expected for LDP mechanisms with both one-time and longitudinal versions (\eg, GRR and L-GRR).
\end{itemize}

\subsection{Main Results} \label{sub:results}

\begin{figure}[!h]
    \centering
    \includegraphics[width=1\linewidth]{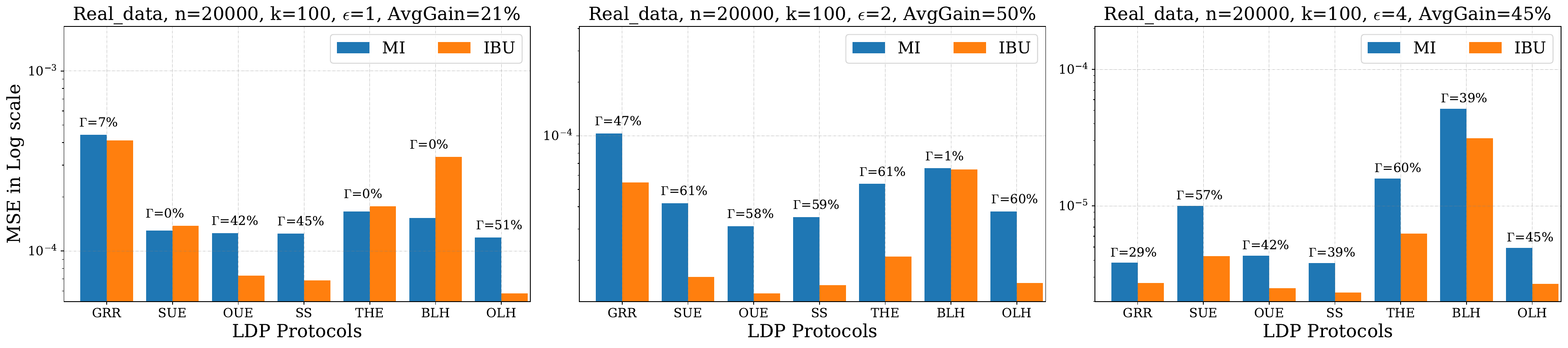}
    \caption{IBU utility gain in \% (\cf{} Eq.~\eqref{eq:utility_gain}) for each one-time pure LDP mechanism with the MSE metric, the real distribution, $n=20000$, $k=100$, and $\epsilon \in \{1, 2, 4\}$.
    AvgGain indicates the average gain considering all LDP mechanisms in the same subplot.}
    \label{fig:IMG_UG_REAL}
\end{figure}

\begin{figure}[ht]
    \centering
    \includegraphics[width=1\linewidth]{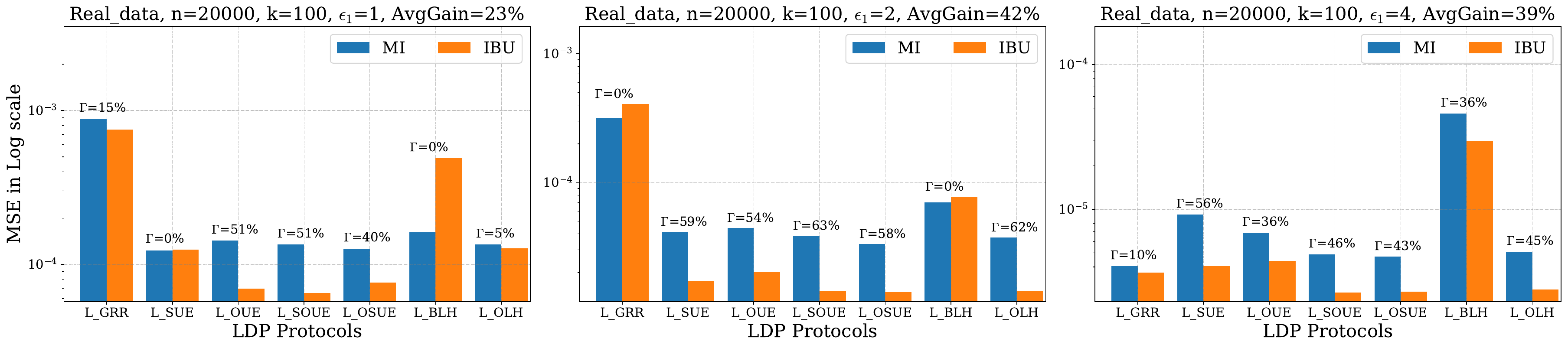}
    \caption{IBU utility gain in \% (\cf{} Eq.~\eqref{eq:utility_gain}) for each longitudinal pure LDP mechanism with the MSE metric, the real distribution, $n=20000$, $k=100$ $\epsPerm \in \{2, 4, 8\}$, and $\epsOne=\frac{\epsPerm}{2}$.
    AvgGain indicates the average gain considering all LDP mechanisms in the same subplot.}
    \label{fig:IMG_UG_REAL_LONG}
\end{figure}

First, considering the real data distribution and the MSE metric, Fig.~\ref{fig:IMG_UG_REAL} (one-time LDP mechanisms) and Fig.~\ref{fig:IMG_UG_REAL_LONG} (longitudinal LDP mechanisms) illustrate the IBU utility gain calculated as in Eq.~\eqref{eq:utility_gain}, by fixing $n=20000$, $k=100$, and varying $\epsilon \in \{1, 2, 4\}$ (resp. $\epsilon=\epsOne$).
These two figures also illustrate the Average Gain (AvgGain) considering all LDP mechanisms within the same subplot.
From Fig.~\ref{fig:IMG_UG_REAL}, one can notice that IBU consistently and considerably outperformed MI for all LDP mechanisms in the medium ($\epsilon=2$) and low privacy regimes ($\epsilon=4$).
For $\epsilon=1$, IBU was only better for GRR, OUE, SS, and OLH, while achieving similar results for SUE and THE, and unsatisfactory results for BLH.
Similar behavior can be noticed on the left-side plot of Fig.~\ref{fig:IMG_UG_REAL_LONG} in which IBU did not improve over MI for L-SUE and L-BLH.
This is because both privacy guarantees are equal for both one-time and longitudinal LDP mechanisms, \ie, $\epsOne=\epsilon$, and thus, similar results will occur, on expectation, when running more iterations for stability.
Overall, for both Fig.~\ref{fig:IMG_UG_REAL} and Fig.~\ref{fig:IMG_UG_REAL_LONG}, the higher AvgGain occurs in the medium privacy regime, followed by low and high privacy regimes, respectively.

\begin{figure}[!ht]
    \centering
    \includegraphics[width=1\linewidth]{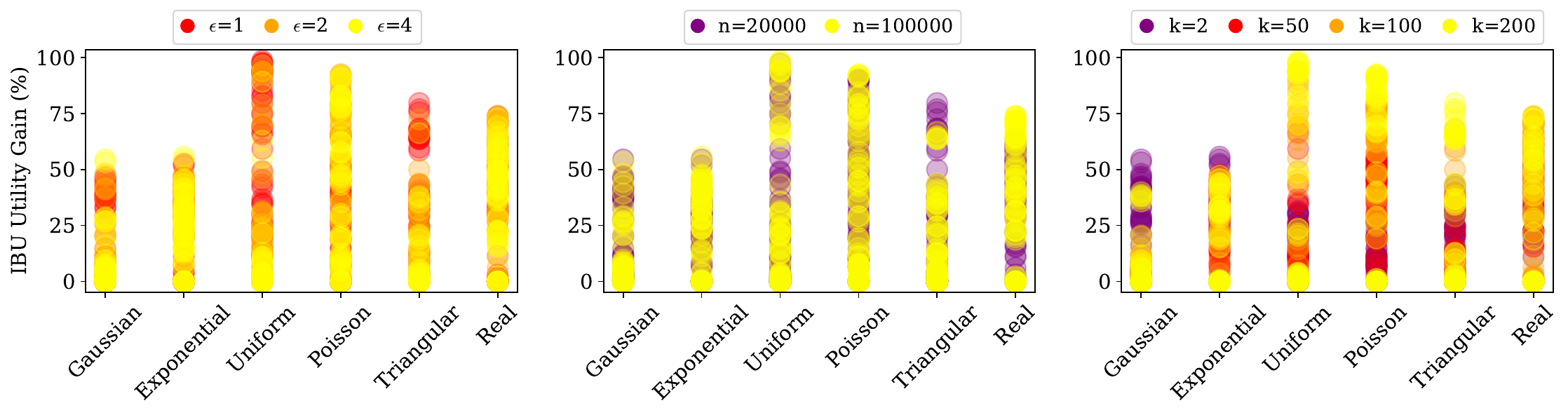}
    \caption{IBU utility gain is in \%. Each subplot shows all gains obtained by distribution type when the privacy level ($\epsilon$), number of users ($n$), and domain size ($k$) are varied, respectively from left to right. We computed all gains with all one-time pure LDP mechanisms considered and the MSE metric.}
    \label{fig:IMG_UG_EPS_N_K}
\end{figure}

\begin{figure}[!ht]
    \centering
    \includegraphics[width=1\linewidth]{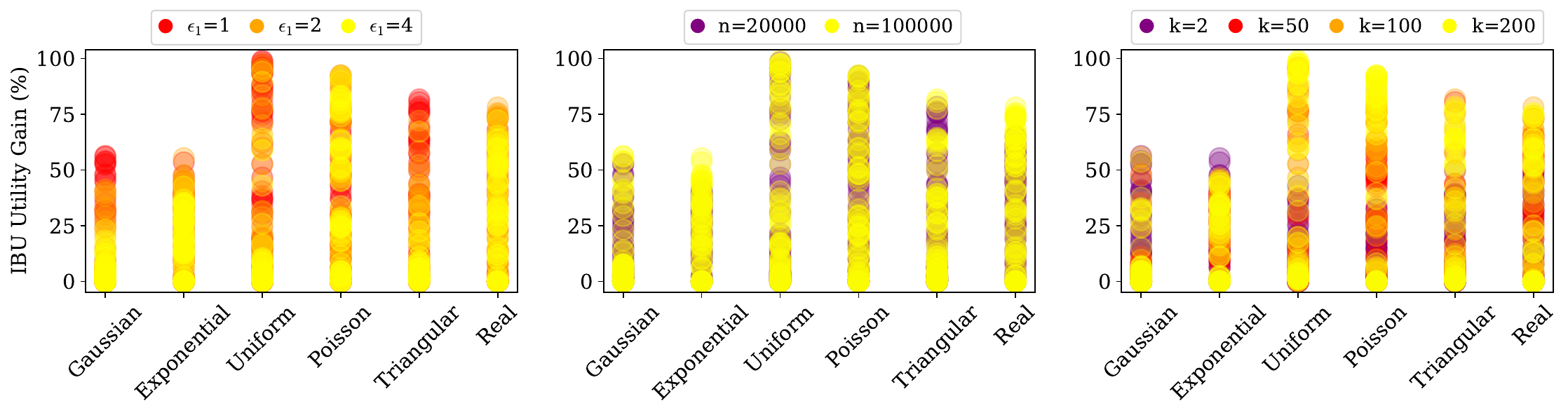}
    \caption{IBU utility gain is in \%. Each subplot shows all gains obtained by distribution type when the privacy level ($\epsilon$), number of users ($n$), and domain size ($k$) are varied, respectively from left to right. We computed all gains with all longitudinal pure LDP mechanisms considered and the MSE metric.}
    \label{fig:IMG_UG_EPS_N_K_LONG}
\end{figure}

Second, in Fig.~\ref{fig:IMG_UG_EPS_N_K} (one-time LDP mechanisms) and Fig.~\ref{fig:IMG_UG_EPS_N_K_LONG} (longitudinal LDP mechanisms), we compare the IBU utility gain in different types of distributions (including real data) by varying $\epsilon$ (left-side plots), $n$ (centered plots), and $k$ (right-side plots). 
In both figures, we observed that the uniform distribution presented the highest gain, while the Gaussian and Exponential distributions obtained the lowest gains. 
This is due to the fact that more values close or equal to zero are generated in Gaussian and Exponential distributions, which makes estimation difficult not only for IBU but also for MI. 
When analyzing the variation of $\epsilon$ in both figures, we noticed that the lower $\epsilon$ (high privacy regime), the higher the IBU gain, which is desirable in practice. 
Furthermore, when examining the variation of the number of users $n$, we noticed that the gains obtained considering all LDP mechanisms for all data distributions present a better performance when $n$ is higher, especially with longitudinal mechanisms. 
This is because longitudinal mechanisms have a higher variance and require more samples to reduce their estimation error~\cite{rappor,Arcolezi2023,Arcolezi2022_allomfree}. 
Last, when analyzing the behavior of the domain size $k$, in both figures, it is shown that IBU outperforms MI, as the domain grows, \ie, IBU supports data with high dimensionality, which better reflects real-world data collections.

Finally, Table~\ref{tab:results_dist_protocols_one_time} (one-time LDP mechanisms) and Table~\ref{tab:results_dist_protocols_long} (longitudinal LDP mechanisms) exhibit the averaged utility gain of IBU for all LDP mechanisms, all data distributions, and both MSE and MAE metrics by considering $k\in \{2,50,100,200\}$, $n\in \{20000,100000\}$, and $\epsilon \in \{1, 2, 4\}$ (resp. $\epsOne=\epsilon$).
From these tables, one can gather the averaged gain for each utility metric (MSE and MAE) by LDP mechanism considering all data distributions (\ie, last row).
For instance, in Table~\ref{tab:results_dist_protocols_one_time}, while the THE mechanism is the one which presented the higher utility gain followed by the SUE mechanism, GRR presented the lowest utility gain.
Similar results can be observed in Table~\ref{tab:results_dist_protocols_long}, \ie, the L-SUE mechanism presented the higher utility gain and L-GRR the lowest one.
In~\cite{ElSalamouny2020}, the authors also remarked that the IBU utility gain for GRR was not substantial.
Indeed, while in Table~\ref{tab:results_dist_protocols_long} L-UE mechanisms showed higher utility gain than L-LH mechanisms, in Table~\ref{tab:results_dist_protocols_one_time}, both UE and LH present similar results.
Additionally, from Table~\ref{tab:results_dist_protocols_one_time} and Table~\ref{tab:results_dist_protocols_long}, one can also gather the averaged gain for each utility metric (MSE and MAE) by data distribution considering all LDP mechanisms (\ie, last two columns).
For example, for both tables, the higher utility gain was observed for the Poisson distribution followed by the real data distribution.
The lower utility gain considering all LDP mechanisms was for the Gaussian distribution followed by the Triangular distribution.

\renewcommand\arraystretch{1.2}
\setlength{\tabcolsep}{1.2pt}
\begin{table}[htb]
    \centering
    \scriptsize
    \begin{tabular}{lcccccccccccccccccc | cc}
    \toprule
    \multirow{2}{*}{Dist.} & \multicolumn{2}{c}{GRR} & \multicolumn{2}{c}{SUE} & \multicolumn{2}{c}{OUE} & \multicolumn{2}{c}{SS} & \multicolumn{2}{c}{THE} &  \multicolumn{2}{c}{BLH} & \multicolumn{2}{c|}{OLH}  & \multicolumn{2}{c}{Avg.} \\ 
    & MSE         & MAE        &  MSE       &   MAE      &   MSE      & MAE        &   MSE      & MAE        &  MSE       &  MAE       & MSE        & MAE        &   MSE      &  \multicolumn{1}{c|}{MAE}   & MSE & MAE    \\\cline{1-17}
    Gauss.    &        1 &        1 &       13 &        7 &       10 &        6 &       3 &       1 &       13 &        7 &       16 &        9 &       11 &        \multicolumn{1}{c|}{7} &        9 &        5 \\
    Exp.      &       16 &       11 &       26 &       15 &       27 &       16 &      19 &      11 &       26 &       15 &       16 &       10 &       27 &       \multicolumn{1}{c|}{16} &       22 &       13 \\
    Unif.     &        0 &        0 &       29 &       21 &       20 &       14 &      14 &      10 &       31 &       22 &       57 &       43 &       18 &       \multicolumn{1}{c|}{12} &       24 &       17 \\
    Poiss.    &       39 &       28 &       41 &       26 &       44 &       28 &      41 &      27 &       41 &       27 &       14 &        6 &       46 &       \multicolumn{1}{c|}{30} &       \textbf{38} &       \textbf{24} \\
    Triang.   &        0 &        0 &       21 &       13 &       15 &        9 &      10 &       6 &       23 &       14 &       36 &       21 &       15 &        \multicolumn{1}{c|}{9} &       17 &       10 \\
    Real      &       31 &       21 &       40 &       23 &       42 &       25 &      34 &      19 &       42 &       25 &       21 &       11 &       44 &       \multicolumn{1}{c|}{27} &       \textbf{36} &       \textbf{21} \\ \hline
    Avg.      &       14 &       10 &       \textbf{28} &       \textbf{17} &       26 &       16 &      20 &      12 &       \textbf{29} &       \textbf{18} &       26 &       16 &       26 &       \multicolumn{1}{c|}{16} &       24 &       15 \\
    \bottomrule
    \end{tabular}
    \caption{Averaged IBU utility gain in \% (\cf{} Eq.~\eqref{eq:utility_gain}) for all one-time pure LDP mechanisms and all data distributions (Dist.), considering $k\in \{2,50,100,200\}$, $n\in \{20000,100000\}$, and $\epsilon \in \{1, 2, 4\}$.
    Results highlighted in \textbf{bold font} represent the two highest utility gains, on average.}
    \label{tab:results_dist_protocols_one_time}
\end{table}

\begin{table}[htb]
    \centering
    \scriptsize
    \begin{tabular}{lcccccccccccccccccc | cc}
    \toprule
    \multirow{2}{*}{Dist.} & \multicolumn{2}{c}{L-GRR} & \multicolumn{2}{c}{L-SUE} & \multicolumn{2}{c}{L-OUE} & \multicolumn{2}{c}{L-SOUE} & \multicolumn{2}{c}{L-OSUE} &  \multicolumn{2}{c}{L-BLH} & \multicolumn{2}{c|}{L-OLH}  & \multicolumn{2}{c}{Avg.} \\ 
    & MSE         & MAE        &  MSE       &   MAE      &   MSE      & MAE        &   MSE      & MAE        &  MSE       &  MAE       & MSE        & MAE        &   MSE      &  \multicolumn{1}{c|}{MAE}   & MSE & MAE    \\\cline{1-17}
    Gauss.    &         14 &          5 &         13 &          8 &          9 &          5 &          10 &           7 &          12 &           7 &          2 &          0 &          7 &          \multicolumn{1}{c|}{4} &        9 &        5 \\
    Exp.      &          4 &          1 &         27 &         16 &         26 &         15 &          27 &          16 &          27 &          16 &          4 &          2 &         20 &         \multicolumn{1}{c|}{12} &       19 &       11 \\
    Unif.     &         36 &         25 &         31 &         22 &         12 &          8 &          16 &          11 &          18 &          13 &         54 &         43 &         21 &         \multicolumn{1}{c|}{16} &       26 &       19 \\
    Poiss.    &          5 &          2 &         43 &         28 &         48 &         32 &          49 &          32 &          44 &          29 &         11 &          6 &         42 &         \multicolumn{1}{c|}{30} &       \textbf{34} &       \textbf{22} \\
    Triang.   &         28 &         17 &         24 &         15 &         11 &          7 &          13 &           9 &          16 &          10 &         26 &         14 &         14 &          \multicolumn{1}{c|}{9} &       18 &       11 \\
    Real.     &          4 &          1 &         43 &         25 &         43 &         27 &          44 &          27 &          43 &          25 &          9 &          4 &         34 &         \multicolumn{1}{c|}{22} &       \textbf{31} &       \textbf{18} \\  \hline
    Avg.      &         15 &          8 &         \textbf{30} &         \textbf{19} &         24 &         15 &          \textbf{26} &          \textbf{17} &          26 &          16 &         17 &         11 &         23 &         \multicolumn{1}{c|}{15} &       23 &       14 \\
    \bottomrule
    \end{tabular}
    \caption{Averaged utility gain in \% of IBU over MI (\cf{} Eq.~\eqref{eq:utility_gain}) for all longitudinal pure LDP mechanisms and all data distributions (Dist.), considering $k\in \{2,50,100,200\}$, $n\in \{20000,100000\}$, $\epsPerm \in \{2, 4, 8\}$, and $\epsOne=\frac{\epsPerm}{2}$. 
    Results highlighted in \textbf{bold font} represent the two highest utility gains, on average.}
    \label{tab:results_dist_protocols_long}
\end{table}

\noindent \textbf{Implementation details.} The \mfreqldpy\ Python package~\cite{Arcolezi2022_multi_freq_ldpy} is function-based and simulates the LDP data collection pipeline of $n$ users and one server. 
Thus, for each solution and/or mechanism, there is always a \textit{client} and an \textit{aggregator} function.
As \mfreqldpy\ had only aggregator functions based on MI, our implementation of IBU contributes with another aggregator function for LDP mechanism in both \texttt{pure\_frequency\_oracle} and \texttt{long\_freq\_est} modules.
For example, the complete code to execute one-time distribution estimation using IBU~\cite{Agrawal2001,Agrawal2005} with data obfuscated through the GRR~\cite{kairouz2016discrete} mechanism is illustrated in Listing~\ref{fig:example_single} with the resulting estimated frequency for a given set of parameters and a randomly generated dataset. 
One can notice that the \texttt{GRR\_Aggregator\_IBU} function receives as input: the set of obfuscated data (rep), the domain size $k$, the privacy guarantee $\epsilon$, the \# iterations (nb\_iter), and a small tolerance value (tol) that works along with an error function (err\_func).
The three last parameters `nb\_iter', `tol', and `err\_func' are stopping criteria for IBU to terminate, and the values in Listing~\ref{fig:example_single} are the default parameters.

\begin{lstlisting}[language=Python, frame=tb, caption={Code snippet for performing distribution estimation using IBU~\cite{Agrawal2001,Agrawal2005} with data obfuscated through the GRR~\cite{kairouz2016discrete} mechanism.},float,label={fig:example_single}]
# Multi-Freq-LDPy functions for GRR protocol
from multi_freq_ldpy.pure_frequency_oracles.GRR import GRR_Client, GRR_Aggregator_IBU

# NumPy library
import numpy as np

# Parameters for simulation
eps = 1 # privacy guarantee
n = int(1e6) # number of users
k = 5 # attribute's domain size

# Simulation dataset following Uniform distribution
dataset = np.random.randint(k, size=n)

# Simulation of client-side data obfuscation
rep = [GRR_Client(user_data, k, eps) for user_data in dataset]

# Simulation of server-side aggregation
GRR_Aggregator_IBU(rep, k, eps, nb_iter=10000, tol=1e-12, err_func="max_abs")
>>> array([0.199, 0.201, 0.199, 0.202, 0.199])
\end{lstlisting}

\section{Related Work}  \label{sec:rel_work}

The literature on the local DP model~\cite{Xiong2020} has largely explored the issue of minimizing the utility loss of LDP mechanisms. 
On the one hand, some works~\cite{tianhao2017,Bassily2015,wang2016mutual,Min2018,kairouz2016discrete,rappor} focused on designed new encoding and perturbation functions, often leading to new privacy-utility trade-offs as well as robustness to privacy attacks~\cite{Arcolezi2023risks}. 
On the other hand, recent research works~\cite{kairouz2016discrete,Wang2020_post_process,elsalamouny2022,ElSalamouny2020,Murakami2018} focused on improving the estimation method on the server side through post-processing techniques for the MI estimator and using IBU.
For instance, the authors in~\cite{kairouz2016discrete} investigated distribution estimation with GRR and SUE by using two post-processing approaches for MI, namely, a method that clips negative elements of $\fest$ to 0 and re-normalizes $\fest$ so that its sum is 1, and a method that projects $\fest$ onto the probability simplex.
Wang \etal~\cite{Wang2020_post_process} studied and proposed, in total 10 post-processing approaches for MI ranging from methods that enforce only non-negativity of elements in $\fest$ or that $\fest$ sums to 1, and other methods that enforce both.
Experiments in~\cite{Wang2020_post_process} were perfomed with the OLH mechanism.
Regarding IBU~\cite{Agrawal2001,Agrawal2005}, recent works have proposed and investigated its performance for discrete distribution estimation~\cite{elsalamouny2022,ElSalamouny2020,Murakami2018} and for joint distribution estimation~\cite{Fanti2016,Ren2018}.
More precisely, ElSalamouny \& Palamidessi~\cite{ElSalamouny2020,elsalamouny2022} proposed a generalization of IBU for personalized LDP, \ie, considering different privacy guarantees $\epsilon_i$ ($i\in \oneTo{n}$) and different LDP mechanisms (\ie, GRR and SUE).

While the aforementioned research works~\cite{kairouz2016discrete,Wang2020_post_process,elsalamouny2022,ElSalamouny2020,Murakami2018} answer interesting questions experimenting only with the GRR~\cite{kairouz2016discrete}, SUE~\cite{rappor}, and OLH~\cite{tianhao2017} mechanisms, we consider in this work fourteen $\epsilon$-LDP mechanisms, \ie, seven for one-time data collection and seven for multiple data collections.
In our analysis we varied the utility metrics (\ie, MSE and MAE), the data distribution (\ie, synthetic data following standard distribution and one real-world data), the number of users $n$, the domain size $k$, and the privacy guarantees $\epsilon$.
Last, we have also open-sourced the IBU implementation into \mfreqldpy~\cite{Arcolezi2022_multi_freq_ldpy}, thereby enabling researchers to easily use and expand upon our results.

\section{Conclusion \& Perspectives} \label{sec:conclusion}

In conclusion, this paper presents an in-depth investigation into the effectiveness of Iterative Bayesian Update (IBU) as a post-processing technique for improving the utility of LDP mechanisms used for private discrete distribution estimation. 
Based on our experiments on both synthetic and real-world data, we compared the performance of IBU to Matrix Inversion (MI), a standard estimation technique. 
We assessed the utility gain of IBU over MI for seven state-of-the-art LDP mechanisms designed for one-time collection~\cite{kairouz2016discrete,Bassily2015,tianhao2017,rappor,wang2016mutual,Min2018} and for seven state-of-the-art longitudinal LDP mechanisms~\cite{rappor,Arcolezi2022_allomfree,Arcolezi2023} designed for multiple data collections. 
On average, both THE~\cite{tianhao2017} and SUE (a.k.a. Basic One-Time RAPPOR)\cite{rappor} mechanisms showed the highest IBU utility gain in our experiments, which involved varying $n$, $k$, $\epsilon$, and the data distribution. 
Regarding longitudinal LDP mechanisms, L-UE mechanisms~\cite{rappor,Arcolezi2022_allomfree} presented higher IBU utility gain, on average, than L-GRR~\cite{Arcolezi2022_allomfree} and L-LH~\cite{Arcolezi2023} mechanisms.
Overall, our results show that IBU can significantly improve the utility of LDP mechanisms for certain data distributions (\eg, Poisson) and specific settings (\cf{} Table~\ref{tab:results_dist_protocols_one_time} and Table~\ref{tab:results_dist_protocols_long}).

Based on the findings of this paper, there are several areas that could be explored for future work. 
Some potential avenues for further research include investigating the performance of IBU on non-pure LDP mechanisms as well as for high-dimensional data, \eg, $k \gg 200$.
Additionally, we plan to investigate different settings for the IBU initialization and stopping criteria, \ie, considering non-uniform initial distributions and different tolerance calculation functions.
Finally, we also aim to implement GIBU (Generalized IBU)~\cite{ElSalamouny2020} and the estimation methods proposed in~\cite{elsalamouny2022} for personalized LDP, into \mfreqldpy~\cite{Arcolezi2022_multi_freq_ldpy}.